\documentclass[pra,10pt,showpacs]{revtex4}
\usepackage{amsmath}    % need for subequations
\usepackage{graphicx}   % need for figures
\usepackage{verbatim}   % useful for program listings
\usepackage{color}      % use if color is used in text
\usepackage{hyperref}   % use for hypertext links, including those to 
\begin{document}

\title{Continuous wave solutions in spinor Bose-Einstein condensates}

\author{Richard S. Tasgal and Y. B. Band}

\affiliation{Departments of Chemistry and Electro-Optics and
the Ilse Katz Center for Nano-Science, \\
Ben-Gurion University of the Negev,
Beer-Sheva 84105,
Israel}

\begin{abstract}
We find analytic continuous wave (cw) solutions for spinor
Bose-Einstein condensates (BECs) in a magnetic field that are more
general than those published to date.  For particles with spin $F=1$
in a homogeneous one-dimensional trap, there exist cw states in which
the chemical potential and wavevectors of the different spin
components are different from each other.  We include linear and
quadratic Zeeman splitting.  Linear Zeeman splitting, if the magnetic
field is constant and uniform, can be mathematically eliminated by a
gauge transformation, but quadratic Zeeman effects modify the cw
solutions in a way similar to non-zero differences in the wavenumbers
between the different spin states.  The solutions are stable fixed
points within the continuous wave framework, and the coherent spin
mixing frequencies are obtained.
\end{abstract}

\date{\today}

% dynamic properties, 03.75.Kk
% multicomponent and spinor condensates, 03.75.Mn
% static properties, 03.75.Hh

\maketitle

\section{Introduction}

Atomic Bose-Einstein condensates (BECs) with nonvanishing total
angular momentum have been experimentally obtained, in which the BECs
contain several spin components, i.e., they are spinor BECs.  Examples
are $H$ (with total spin $F=0$ and $F=1$) \cite{Fried.1998}, $^7$Li
($F=1$, $F=2$; note that the scattering length is negative, making the
BEC self-attractive) \cite{Bradley.1995,Bradley.1997}, $^{23}$Na
($F=1$, $F=2$) \cite{Simkin.1999}, $^{39}$K ($F=1$, $F=2$; scattering
length has a small negative value) \cite{Roati.2007}, $^{41}$K ($F=1$,
$F=2$) \cite{Modugno.2001}, $^{52}$Cr ($F=3$; with a large magnetic
moment) \cite{Griesmaier.2005, Beaufils.2008}, $^{85}$Rb ($F= 2$,
$F=3$; negative scattering length), $^{87}$Rb ($F=1$, $F=2$)
\cite{Barrett.2001, Chang.2004, Sadler.2006, Leslie.2009}, $^{133}$Cs
($F=3$, $F=4$) \cite{Weber.2003}, $^{164}$Dy ($F=8$) \cite{Lu.2011},
and $^{168}$Er ($F=6$) \cite{Aikawa.2012}.
% Spinor BECs have been produced with atoms in excited electronic
% states that have non-zero spin even though the ground state is
% scalar ($F=0$):  $^4$He, $^{40}$Ca \cite{Kraft.2009},
% $^{84}$Sr \cite{Stellmer.2009}, $^{86}$Sr \cite{},
% $^{88}$Sr \cite{}, $^{170}$Yb (I=0) \cite{Fukuhara.2007},
% and $^{174}$Yb \cite{Takasu.2003},
When a BEC is held in a magnetic trap, only spin components of one
sign are trapped, often leaving just one spin orientation.  An
\textit{optical} trap, however, may hold all spin components of a
given hyperfine state.  In this case, the spinor character is
important, and a scalar model of a BEC is insufficient.  Optical traps
for spinor BECs are now common, and there is work on creating BECs
using only optical traps \cite{Barrett.2001, Beaufils.2008}.  Optical
traps have also been created in the form of a closed ring
\cite{Gupta.2005, Olson.2007, Lesanovsky.2007, Henderson.2009,
Ryu.2007}.

The spinor properties of BECs have been the focus of recent
theoretical and experimental research \cite{Ueda.2012,
KawaguchiUeda.2012}, especially $^{23}$Na, $^{87}$Rb, and $^{52}$Cr.
The spinor properties of a BEC can critically affect the dynamics: The
spinor character of BECs underlies the formation of domain walls
(transition regions, which may be stable or unstable, between distinct
spin domains) and vortices \cite{Sadler.2006, Saito.2007,
Lamacraft.2007}.  Oscillatory coherent spin mixing occurs in spinor
BECs \cite{Zhang.2005, Gerbier.2006}.  Spinor BECs are subject to
modulational (Benjamin-Feir) instabilities \cite{Robins.2001,
Konotop.2002, Li.2005, KawaguchiUeda.2012} even when the nonlinearity is repulsive,
whereas scalar BECs are not.  Spinor BECs support a variety of soliton solutions that are not
found in scalar BECs \cite{IedaMiyakawaWadati.2004,
WadatiTsuchida.2006, UchiyamaIedaWadati.2006,
SzankowskiTrippenbachInfeldRowlands.2010,
SzankowskiTrippenbachInfeldRowlands.2011}.
% 
% SzankowskiTrippenbachInfeld.2011
% SzankowskiTrippenbachInfeld.2012

Here we study continuous wave (cw) [plane wave] solutions of $F=1$
spinor BEC condensates.  Section~\ref{Sec:SpinorBEC.model} introduces
the model, which includes spin-dependent and spin-independent
mean-field effects, and linear and quadratic Zeeman effects (but does
not include spin-dipolar effects, which can be important, e.g., in
$^{52}$Cr).  Section~\ref{Sec:SpinorBEC.cw} derives the most general
possible cw solutions for $F=1$ spinor BECs on a homogeneous
background with a homogeneous magnetic field.  This section gives,
even before inclusion of the linear and quadratic Zeeman effects, more
general families of cw solutions than have heretofore been published.
Section~\ref{Sec:SpinorBEC.conclusions} contains a summary and
conclusions.

\section{Quantitative model for spinor BECs with magnetic fields}
\label{Sec:SpinorBEC.model} 

The Hamiltonian density for an $F=1$ spinor BEC with linear and
quadratic Zeeman effect (and without significant spin-dipolar coupling) is
\cite{OhmiMachida.1998, Ho.1998}
\begin{eqnarray}
\mathcal{H}
& = &
\frac{\hbar^2}{2 m} \mathbf{\nabla} \Phi_a^\dagger \cdot \mathbf{\nabla} 
\Phi_a + \frac{c_0}{2} \Phi_a^\dagger \Phi_b^\dagger \Phi_b \Phi_a
+ \frac{c_2}{2} \Phi_a^\dagger \Phi_{a'}^\dagger \mathbf{F}_{ab} \cdot \mathbf{F}_{a'b'} \Phi_{b'} \Phi_b
- p B \Phi_a^\dagger \mathbf{F}_{ab}^z \Phi_b
+ q B^2  \Phi_a^\dagger (\mathbf{F}_{ab}^z)^2 \Phi_b,
\end{eqnarray}
where $\mathbf{\Phi} = (\phi_1,\phi_0,\phi_{-1})^\mathrm{t}$ is a
vector with the amplitudes of the $M_F =1$ component (spin in the same
direction as the magnetic field), $M_F = 0$, and $M_F = -1$; $m$ is
the mass of the atom, $c_0$ and $c_2$ are the coefficients of the
spin-independent and spin-dependent parts of the mean-field; $c_0$
gives rise to self-phase modulation and $c_2$ is the spin-dependent
mean-field coefficient which also gives rise to phase modulation, and
parametric nonlinearity.  $\mathbf{F}$ is a vector in which each
component is a 3$\times$3 spin-1 matrix, i.e., the dimensionless spin
vector $\mathbf{F}$ has components,
\begin{equation}   \label{Eq:5.AMrl.j.eq.1}
  F_x  = \frac{1}{\sqrt{2}} \left( \! \!  
  {\begin{array}{ccc}
   0 & 1 & 0  \\
   1 & 0 & 1  \\
   0 & 1 & 0 
 \end{array} }  \! \! \right), \quad
 F_y  = \frac{1}{\sqrt{2}} 
 \left( \! \!  {\begin{array}{ccc}
   0 & { - i} & 0 \\
   i & 0 & {-i}  \\
   0 & i & 0
 \end{array} }  \! \! \right), \quad
 F_z  = 
 \left(  \! \! {\begin{array}{ccc}
   1 & 0 & 0  \\
   0 & 0 & 0  \\
   0 & 0 & {-1}
 \end{array} }  \! \! \right) .
\end{equation}
The magnetic field ${\bf B}$ is taken to be constant and uniform and
in the $z$-direction, ${\bf B} = B \hat {\mathbf z}$, and $p$ and $q$
are linear and quadratic Zeeman coefficients \cite{Saito.2007,
Lamacraft.2007, Zhang.2005}.  If the BEC is confined to one dimension
(by a strong optical trap in the transverse directions), the
Hamiltonian gives the following governing equations for the BEC:
\begin{subequations}
\label{Eqs:SpinorBEC}
\begin{eqnarray}
i \hbar \frac{\partial}{\partial t} \phi_1 
& = & -\frac{\hbar^2}{2 m} \frac{\partial^2}{\partial z^2} \phi_1
           + c_0 \left( |\phi_1|^2 + |\phi_0|^2 + |\phi_{-1}|^2 \right) \phi_1
             \nonumber \\
& & + c_2 \left[ \left( |\phi_1|^2 + |\phi_0|^2 - |\phi_{-1}|^2 \right) \phi_1 
	                  + \phi_0^2 \phi_{-1}^*
	             \right]
        + (-p B + q B^2) \phi_1 \, ,
      \label{Eq:phi+1} \\
i \hbar \frac{\partial}{\partial t} \phi_0 & = &
           - \frac{\hbar^2}{2 m} \frac{\partial^2}{\partial z^2} \phi_0
           + c_0 \left( |\phi_1|^2 + |\phi_0|^2 + |\phi_{-1}|^2 \right) \phi_0 
	   \nonumber \\
& &           
           + c_2 \left[ \left( |\phi_1|^2 + |\phi_{-1}|^2 \right) \phi_0
                            + 2 \phi_1 \phi_0^* \phi_{-1}
                      \right] \, ,
      \label{Eq:phi0} \\
i \hbar \frac{\partial}{\partial t} \phi_{-1} 
& = & -\frac{\hbar^2}{2 m} \frac{\partial^2}{\partial z^2} \phi_{-1}
           + c_0 \left( |\phi_1|^2 + |\phi_0|^2 + |\phi_{-1}|^2 \right) 
	   \phi_{-1} \nonumber \\
& &  + c_2 \left[ \left( -|\phi_1|^2 + |\phi_0|^2 + |\phi_{-1}|^2 \right) 
                            \phi_{-1} + \phi_0^2 \phi_1^*
                      \right]
       + (p B + q B^2) \phi_{-1} \, .
      \label{Eq:phi-1}
\end{eqnarray}
\end{subequations}
Time and space are denoted by $t$ and $z$, respectively.  The system
has Galilean invariance.  Materials with negative $c_2$ are
ferromagnetic, since with $c_2<0$, at a given particle density and at
zero magnetic field, the energy density is lower when the local BEC is
in a pure spin state $M_F = 1$ or $M_F = -1$, than in a state composed
of a mixture of spins $M_F = 1$ and $M_F = -1$.  The BEC prefers to
have a non-zero net spin locally.  Materials with positive $c_2$ are
antiferromagnetic, or polar, since with $c_2<0$, at a given particle
density and without a magnetic field, the energy density is lower when
the local BEC is 50\% $M_F = 1$ and 50\% $M_F = -1$.  The BEC prefers
to be in a state where opposite spins balance each other exactly and,
cancel out in the total magnetic moment.  The nonlinear coefficients,
which are proportional to the s-wave scattering lengths $a_0$, $a_2$
for the total spin $F=0$ and $F=2$ channels, $g_0 = (4\pi\hbar^2/m)
a_0$, $g_2 = (4\pi\hbar^2/m) a_2$, with the nonlinear coefficients in
the governing equations above given by $c_0 = (g_0 + 2 g_2) / 3$, $c_2
= -(g_0 - g_2) / 3$.  If the BEC is confined to one dimension, the
values of the nonlinear coefficients are modified as discussed in
Refs.~\cite{Olshanii.1998, Bergeman.2003}.
% The nonlinear coefficients are functions of the atomic masses ($m$),
% the s-wave scattering lengths $a_0$, $a_2$ for the spin-0 and spin-2
% channels, and, if the BEC is confined to one dimension, the
% transverse oscillator lengths ($a_{\perp 0} = \sqrt{ \hbar / m
% \omega_{\perp 0}}$, $a_{\perp 2} = \sqrt{\hbar / m \omega_{\perp
% 2}}$), where $\omega_{\perp 0}$, $\omega_{\perp 2}$ are the
% transverse trap frequencies \cite{Olshanii.1998, Bergeman.2003}.  In
% bulk, where $|\phi_j|^2$ represents particles per volume,
% $g_0^{3D} = (4\pi\hbar^2/m) a_0$,
% $g_2^{3D} = (4\pi\hbar^2/m) a_2$,
% and in 1-dimension, where $|\phi_j|^2$ is particles per unit length,
% $g_0^{1D} = (4\pi\hbar^2/m) a_0 / 
% [(\pi a_{\perp 0}^2) (1 + \zeta(1/2) a_0 / a_{\perp 0})]$,
% $g_2^{1D} = (4\pi\hbar^2/m) a_2 /[(\pi a_{\perp 2}^2) (1 + \zeta(1/2) 
% a_2 / a_{\perp 2})]$,
% where $\zeta(1/2) \approx 1.4$
% \cite{Olshanii.1998, Bergeman.2003}. 
% In both instances, the nonlinear coefficients of the governing equations 
% are
% $c_0 = (g_0 + 2 g_2) / 3$,
% $c_2 = -(g_0 - g_2) / 3$.
%
For $^{87}$Rb, the scatterings lengths are $a_0 = 101.8\,a_B$ and $a_2
= a_0 - 1.45\,a_B$, where $a_B$ is the Bohr radius \cite{Chang.2005,
KlausenBohnGreene.2001, KempenKokkelmansHeinzenVerhaar.2002}.  This
gives $^{87}$Rb a negative $c_2$, which makes it ferromagnetic.  For
$^{23}$Na, the scattering lengths have been measured to be $a_0 =
50.0\,a_B$, $a_2 = a_0 + 5.0\,a_B$.  $^{23}$Na has a positive $c_2$,
so it is antiferromagnetic, or ``polar.''  The masses of the two atoms
are $m_{\mathrm{Rb}} = 1.4192 \times 10^{-22}$g, $m_{\mathrm{Na}} =
3.817 \times 10^{-23}$g.  In bulk, these yield $c_0(\mathrm{Rb})= 5.26
\times 10^{-38}$\,erg\,cm$^3$, $c_2(\mathrm{Rb}) = -2.52 \times
10^{-40}$\,erg\,cm$^3$ and $c_0(\mathrm{Na}) = 10.33 \times
10^{-38}$\,erg\,cm$^3$, $c_2(\mathrm{Na}) = 3.23 \times
10^{-39}$\,erg\,cm$^3$; the rations $c_2/c_0$ are $-0.0048$ for
$^{87}$Rb and $0.0313$ for $^{23}$Na.  The quadratic Zeeman
coefficient is $q=h \times 575$\,Hz/G$^2$ for $^{87}$Rb,
% 
% the magnetic field induces a quadratic Zeeman shift of $575.14\times
% 10^{8}$HzT-2.  S. Bize, Y. Sortais, M. S. Santos, C. Mandache, A.
% Clairon, and C. Salomon, ``High-accuracy measurement of the
% $^{87}\rm Rb$ ground-state hyperfine splitting in an atomic
% fountain,'' Europhys.  Lett.  \textbf{45}, 558-564 (1999).  h $575$
% Hz/G^2
% 
and $q=h \times 70$\,Hz/G$^2$ for $^{23}$Na.
Equations~(\ref{Eqs:SpinorBEC}) are integrable when $c_2=0$ (in which
case the system is a set of generalized Manakov equations
\cite{Nakkeeran.1998, Manakov.1973}) or $c_2=c_0$
\cite{IedaMiyakawaWadati.2004, WadatiTsuchida.2006, IedaWadati.2007}.

The linear Zeeman splitting can be eliminated from the governing
equations~(\ref{Eqs:SpinorBEC}) by the change of variables (i.e., the
gauge transformation) $\phi_1 \equiv \psi_1 \exp[i (p B / \hbar) t]$,
$\phi_0 \equiv \psi_0$, $\phi_{-1} \equiv \psi_{-1} \exp[-i (p B /
\hbar) t]$.  Let us take this as done, but retain the same variable
names as previously, to avoid excessive complicated notation.  Because
the linear Zeeman splitting can be eliminated by this gauge
transformation, the analysis of the nil linear Zeeman splitting,
$p=0$, can equally well describe the case with non-zero linear Zeeman
splitting.  Generally, analyses should either make use of this gauge
transformation or not force all the spin components to have the
same frequency (chemical potential), or one may erroneously impose
restrictions that do not exist in the physics.  Quadratic Zeeman
splitting cannot be eliminated by a change of variables.

The analysis below is in terms of infinite continuous (plane) waves.
It applies equally to waves with periodic boundary conditions, either
physical ones due to the BECs existing in a circular trap
\cite{Gupta.2005, Olson.2007, Lesanovsky.2007, Henderson.2009,
Ryu.2007}, or theoretical ones due to the need to limit the numerical
domain.  Periodic boundary conditions quantize the wavenumbers.

The governing equations~(\ref{Eqs:SpinorBEC}) can be
non-dimensionalized by the change of variables
\begin{subequations}
\begin{eqnarray}
t' & = & t / t_d , \\
z' & = & z / z_d = z / \sqrt{\hbar t_d / m} , \\
\phi_j' & = & \phi_j / \phi_d = \phi_j / \sqrt{\hbar / (c_0 t_d)} .
\label{Eq:nondimensionalize}
\end{eqnarray}
\end{subequations}
In dimensionless variables~(\ref{Eq:nondimensionalize}), the
governing equations~(\ref{Eqs:SpinorBEC}) take the same form but with
$\hbar \rightarrow 1$, $m \rightarrow 1$, $c_0 \rightarrow 1$, and
$c_2 \rightarrow c_2/c_0$.  The physical frequencies and wavenumbers
are equal to the dimensionless quantities times $t_d^{-1}$ and
$z_d^{-1} = (\sqrt{\hbar t_d / m})^{-1}$, respectively.  The physical
amplitudes of the BEC are the dimensionless ones times $\sqrt{\hbar /
(c_0 t_d)}$.  The physical energy is equal to the dimensionless energy
multiplied by $\hbar / t_d$, and the physical energy is the
dimensionless quantity times $\hbar^{-1/2} m^{3/2} t_d^{-5/2}$.  We
will use the dimensionless variables in the figures in order to
emphasize the generality, but retain the dimensions in the body of the
text.

The Lagrangian density that gives Eqs.~(\ref{Eqs:SpinorBEC}) is 
\begin{eqnarray}
\label{Spinor.Lagrangian}
\mathcal{L} 
& = &
\frac{i \hbar}{2} (\phi_1^* \phi_{1,t} - \phi_1 \phi_{1,t}^* + 
\phi_0^* \phi_{0,t} - \phi_0 \phi_{0,t}^* + \phi_{-1}^* \phi_{-1,t} 
- \phi_{-1} \phi_{-1,t}^*) \nonumber \\
& - & 
\frac{\hbar^2}{2 m} (|\phi_{1,z}|^2 + |\phi_{0,z}|^2 + 
|\phi_{-1,z}|^2) % \nonumber \\
- % & - &
\frac{c_0}{2} \left( |\phi_1|^2 + |\phi_0|^2 + |\phi_{-1}|^2 \right)^2
\nonumber \\
& - &
c_2
\left[
  \frac{1}{2} \left( |\phi_1|^2 - |\phi_{-1}|^2 \right)^2
  + |\phi_0|^2 \left( |\phi_1|^2 + |\phi_{-1}|^2 \right)
  + \phi_1^* \phi_0^2 \phi_{-1}^* + \phi_1 \phi_0^{*2} \phi_{-1}
\right]
- q B^2 (|\phi_1|^2 + |\phi_{-1}|^2) \, ,
% & - & (-p B + q B^2) |\phi_1|^2
 %        - (p B + q B^2) |\phi_{-1}|^2 ,
\end{eqnarray}
and the Hamiltonian density is
\begin{eqnarray}
\label{Spinor.Hamiltonian}
\mathcal{H} 
& = &
\frac{\hbar^2}{2 m} (|\phi_{1,z}|^2 + |\phi_{0,z}|^2 + |\phi_{-1,z}|^2) % \nonumber \\
+ % &  + &
\frac{c_0}{2}
\left( |\phi_1|^2 + |\phi_0|^2 + |\phi_{-1}|^2 \right)^2
\nonumber \\
& + &
c_2
\left[
  \frac{1}{2} \left( |\phi_1|^2 - |\phi_{-1}|^2 \right)^2
  + |\phi_0|^2 \left( |\phi_1|^2 + |\phi_{-1}|^2 \right)
  + \phi_1^* \phi_0^2 \phi_{-1}^* + \phi_1 \phi_0^{*2} \phi_{-1}
\right]
+ q B^2 (|\phi_1|^2 + |\phi_{-1}|^2) \, .
% & + & (-p B + q B^2) |\phi_1|^2
%     + (p B + q B^2) |\phi_{-1}|^2 .
\end{eqnarray}

%This gives dimensionless governing equations
%\begin{subequations}
%\label{Eqs:SpinorBEC.dimensionless}
%\begin{eqnarray}
%i \frac{\partial}{\partial t'} \phi_1' & = &
%           - \frac{1}{2} \frac{\partial^2}{\partial z'^2} \phi_1'
%           + \left( |\phi_1'|^2 + |\phi_0'|^2 + |\phi_{-1}'|^2 \right) \phi_1'
%           + \frac{c_2}{c_0}
%              \left[ \left( |\phi_1'|^2 + |\phi_0'|^2 - |\phi_{-1}'|^2 \right) 
%              \phi_1'  + \phi_0'^2 \phi_{-1}'^*
%              \right] \, ,
%      \label{Eq:phi+1.dimensionless} \\
%i \frac{\partial}{\partial t'} \phi_0' & = &
%           - \frac{1}{2} \frac{\partial^2}{\partial z'^2} \phi_0'
%           + \left( |\phi_1'|^2 + |\phi_0'|^2 + |\phi_{-1}'|^2 \right) \phi_0'
%           + \frac{c_2}{c_0}
%               \left[ \left( |\phi_1'|^2 + |\phi_{-1}'|^2 \right) \phi_0'
%                     + 2 \phi_1' \phi_0'^* \phi_{-1}'
%              \right] \, ,
%      \label{Eq:phi0.dimensionless} \\
%i \frac{\partial}{\partial t'} \phi_{-1}' & = &
%           - \frac{1}{2} \frac{\partial^2}{\partial z'^2} \phi_{-1}'
%           + \left( |\phi_1'|^2 + |\phi_0'|^2 + |\phi_{-1}'|^2 \right) 
%           \phi_{-1}' + \frac{c_2}{c_0}
%              \left[ \left( -|\phi_1'|^2 + |\phi_0'|^2 + |\phi_{-1}'|^2 \right) 
%              \phi_{-1}' + \phi_0'^2 \phi_1'^*
%              \right] \, .
%      \label{Eq:phi-1.dimensionless}
%\end{eqnarray}
%\end{subequations}

If any two of the three spin fields are zero, then the remaining field
is governed by a simple nonlinear Schr\"{o}dinger equation, which is
completely integrable \cite{ZakharovShabat.1971, ZakharovShabat.1973}.
If the $M_F = 0$ field is nil ($\phi_0=0$), then the $M_F = \pm 1$
fields ($\phi_1$, $\phi_{-1}$) are governed by a pair of coupled
nonlinear Schr\"{o}dinger equations,
\begin{subequations}
\label{CNLS:SpinorBEC}
\begin{eqnarray}
i \hbar \frac{\partial}{\partial t} \phi_1 & = &
           - \frac{\hbar^2}{2 m} \frac{\partial^2}{\partial z^2} \phi_1
           + \left[ (c_0 + c_2) |\phi_1|^2 + (c_0 - c_2) |\phi_{-1}|^2 \right] 
	   \phi_1 \, ,
      \label{CNLS:phi+1} \\
i \hbar \frac{\partial}{\partial t} \phi_{-1} & = &
           - \frac{\hbar^2}{2 m} \frac{\partial^2}{\partial z^2} \phi_{-1}
           + \left[ (c_0 - c_2) |\phi_1|^2 + (c_0 + c_2) |\phi_{-1}|^2 \right] 
	   \phi_{-1} \, .
      \label{CNLS:phi-1}
\end{eqnarray}
\end{subequations}
The coupled nonlinear Schr\"{o}dinger equations have been intensely
studied (see, e.g., Ref.~\cite{Yang.2010}).  This case is completely
integrable if and only if $c_2=0$, $c_0=c_2$, or $c_0=0$
\cite{ZakharovSchulman.1982}.  Depending on the coefficients, there
may be bright soliton solutions \cite{Manakov.1973, Agrawal.2001},
dark soliton solutions \cite{KivsharTuritsyn.1993}, bright-dark
soliton solutions \cite{RadhakrishnanLakshmanan.1995,
RadhakrishnanLakshmananDaniel.1995}, and domain wall solutions
\cite{HaeltermanSheppard.1994, Malomed.1994}.  We will concentrate
less on this limit, and more on aspects of spinor BECs that do not
overlap with the thoroughly studied coupled nonlinear Schr\"{o}dinger
equations.

\section{Continuous wave solutions}
\label{Sec:SpinorBEC.cw}

Continuous waves (cws) are the simplest shapes, so cw solutions are
where analysis of the system should begin.  The \textit{most general}
cw is
\begin{subequations}
\label{cw.ansatz}
\begin{eqnarray}
\phi_1 & = & A_1 \exp[i (\theta_1 + k_1 z - \omega_1 t)]  \, ,
      \label{cw.ansatz.phi+1} \\
\phi_0 & = & A_0 \exp[i (\theta_0 + k_0 z - \omega_0 t)] \, ,
       \label{cw.ansatz.phi0} \\
\phi_{-1} & = & A_{-1} \exp[i (\theta_{-1} + k_{-1} z - \omega_{-1} t)] \, ,
      \label{cw.ansatz.phi-1}
\end{eqnarray}
\end{subequations}
where the parameters are real-valued and, without loss of generality,
$A_1$, $A_0$, $A_{-1}$ are positive definite.  Note that this is more
general than the cw ansatz in
Ref.~\cite{WadatiTsuchida.2006}, in that the frequencies and
wavenumbers need not all be the same.  The analysis herein shows that
there exist a wider range of cw solutions than, for example, in
Ref.~\cite{WadatiTsuchida.2006}.  Put the trial
function~(\ref{cw.ansatz}) into Eqs.~(\ref{Eqs:SpinorBEC}).

If $c_2 = 0$, then Eqs.~(\ref{Eqs:SpinorBEC}) are a generalized
Manakov system \cite{Manakov.1973}, but with three fields rather than
two.  This limiting case is completely integrable \cite{Nakkeeran.1998}.
There are cw solutions for every value of the amplitude ($A_j$),
wavenumber ($k_j$), and phase ($\theta_j$). 
The frequencies of the fields are
\begin{equation}
    \hbar \omega_j = \frac{ \hbar^2 k_j^2}{2 m} + c_0 (A_1^2 + A_0^2 + A_{-1}^2) \, .
\end{equation} 

\begin{figure}
\centering
% \begin{subfigure}[b]{8.6 cm}
% \centering
\includegraphics[width=8.6 cm]{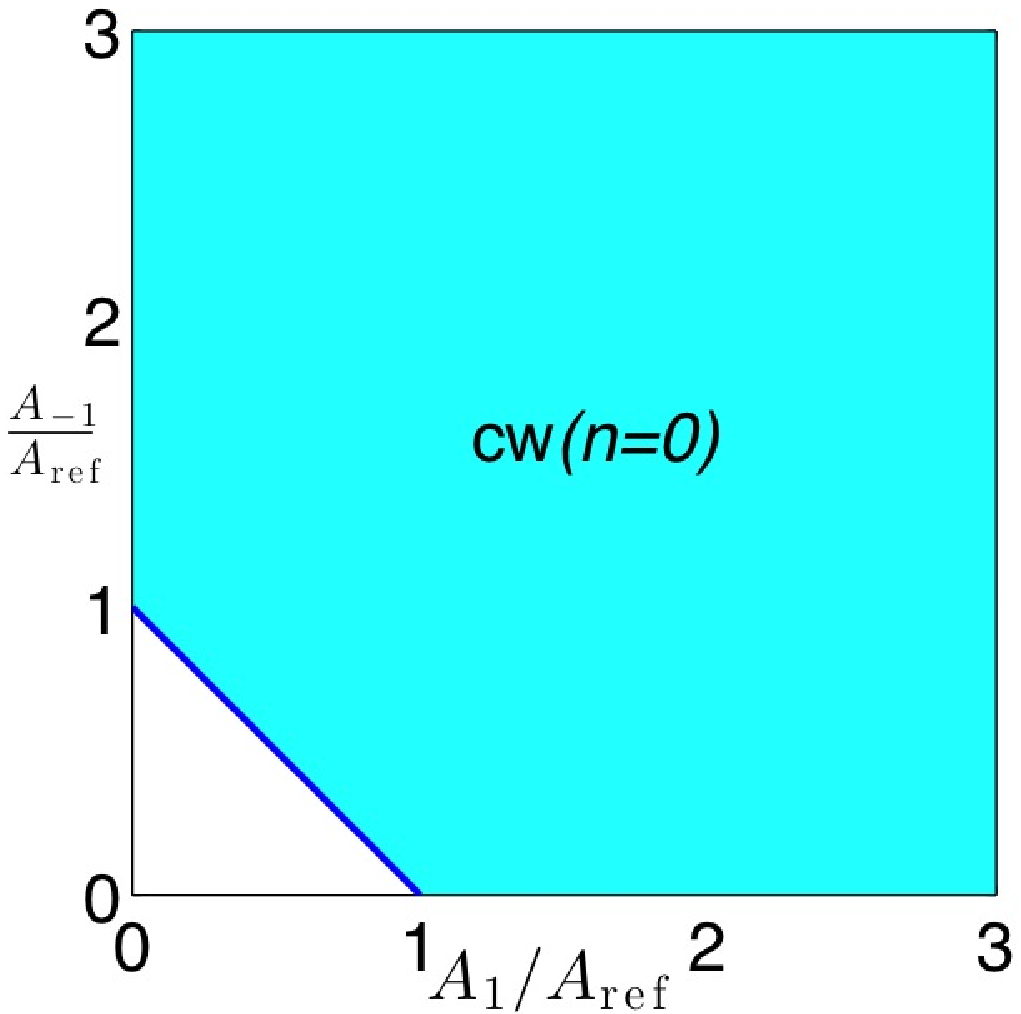}
% \caption{CW($n=0$)}
% \label{Fig.CW.n0.exist}
% \end{subfigure}~%
% \begin{subfigure}[b]{8.6 cm}
% \centering
\includegraphics[width=8.6 cm]{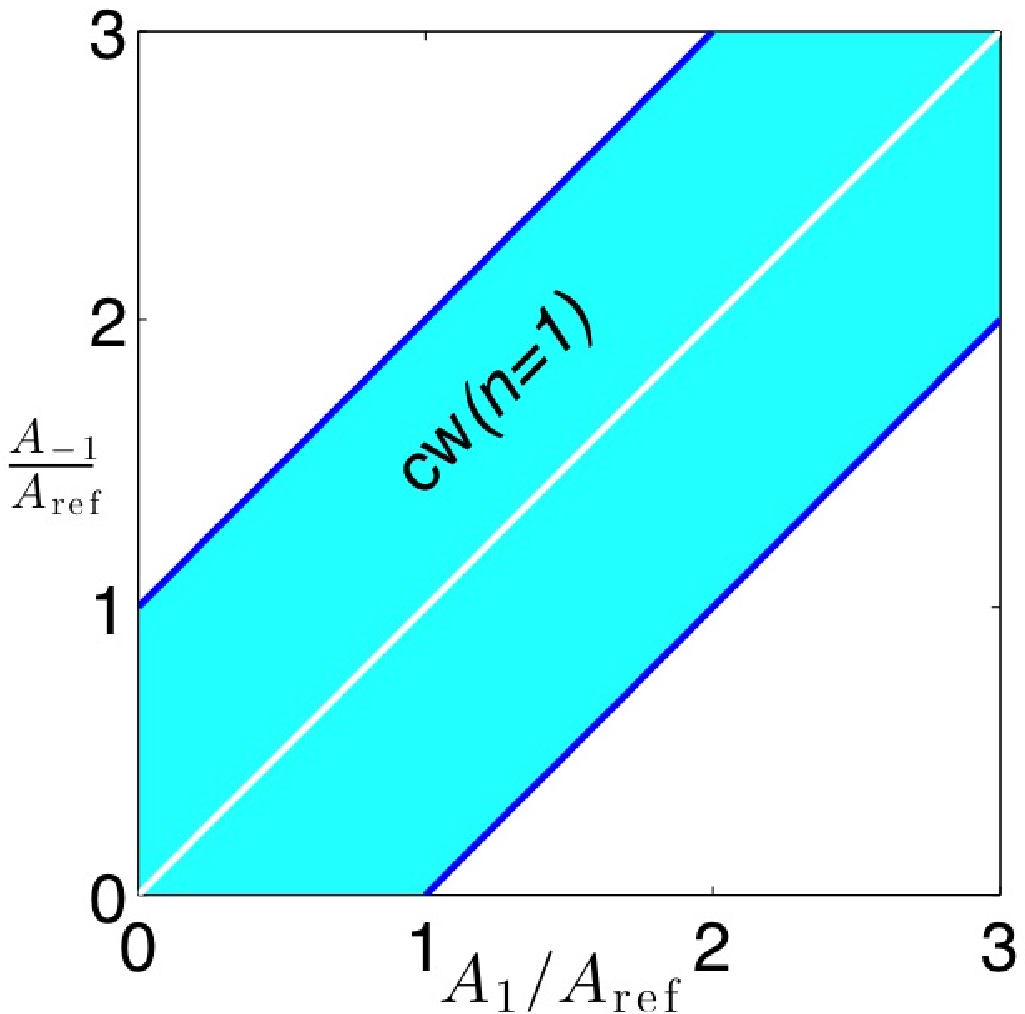}
% \caption{CW($n=1$)}
% \label{Fig.CW.n1.exist}
% \end{subfigure}
\caption{(Color online) Existence ranges of the cw solutions in spinor
BECs in the generalized antiferromagnetic case ($A_\mathrm{ref} \equiv
[ \hbar^2 (k_1 - k_{-1})^2 / 8 m + q B^2] / c_2 \geq 0$; note that in the most familiar cw solutions,
with $k_1=k_0=k_{-1}$ and zero quadratic Zeeman splitting, $A_\mathrm{ref}=0$;
in this limit, the $n=0$ cw solutions have $A_0=\sqrt{2 A_1 A_{-1}}$
and exist for all values of the particle densities $M_F=\pm 1$,\
and the $n=1$-type cw does not exist). 
Part~(a) shows cws with relative phase of the $M_F = 0$ field corresponding to
even $n$ in the trial function Eq.~(\ref{cw.ansatz}), with $\theta_1 +
\theta_{-1} - 2\theta_0 = n \pi$.  Part~(b) shows the cws with
relative phase designated by odd $n$.  For odd $n$, the particle
density of the $M_F = 0$ field approaches infinity as the densities of
$M_F = 1$ and $M_F = -1$ fields approach equality.}
\label{Fig.CWs.c2positive}
\end{figure}

If $c_2 \neq 0$, then the parametric term requires a relation between
the phases of the three fields,
\begin{subequations}
\label{cw.ansatz.phases.0}
\begin{eqnarray}
k_0 & = & \frac{1}{2} (k_1 + k_{-1}) \, , \label{cw.ansatz.k.0} \\
\omega_0 & = & \frac{1}{2} (\omega_1 + \omega_{-1}) \, , 
\label{cw.ansatz.omega.0} \\
\theta_0 & = & \frac{1}{2} (\theta_1 + \theta_{-1} + n \pi) \, , 
\label{cw.ansatz.theta.0} 
\end{eqnarray}
\end{subequations}
where $n$ is an integer.
The equations for the magnitudes of the fields are
\begin{subequations}
\label{cw.magnitudes}
\begin{eqnarray}
\hbar \omega_1& = & \frac{\hbar^2 k_1^2}{2 m}
           + c_0 \left( A_1^2 + A_0^2 + A_{-1}^2 \right)
           + c_2 \left[ A_1^2 + A_0^2 - A_{-1}^2 + (-1)^n A_0^2 A_{-1} / 
	   A_1 \right]
	  + q B^2 A_1 \, ,
% 	  + (-p B + q B^2) A_1 \, ,
      \label{cw.A+1} \\
\hbar \omega_0 & = & \frac{\hbar^2 k_0^2}{2 m}
           + c_0 \left( A_1^2 + A_0^2 + A_{-1}^2 \right)
           + c_2 \left[ A_1^2 + A_{-1}^2 + 2 (-1)^n A_1 A_{-1} \right] \, ,
      \label{cw.A0} \\
\hbar \omega_{-1} & = & \frac{\hbar^2 k_{-1}^2}{2 m}
           + c_0 \left( A_1^2 + A_0^2 + A_{-1}^2 \right)
           + c_2 \left[ -A_1^2 + A_0^2 + A_{-1}^2 + (-1)^n A_0^2 A_1 / 
	   A_{-1} \right] 
	  + q B^2 A_{-1} \, .
% 	  + (p B + q B^2) A_{-1} \, .
	   \label{cw.A-1}
\end{eqnarray}
\end{subequations}
Equations~(\ref{cw.magnitudes}) and~(\ref{cw.ansatz.omega.0}) give a
formula for the magnitude $A_0$ of mode $\phi_0$,
\begin{equation}
\label{cw.A0.solution}
A_0^2 = 2 (-1)^n A_1 A_{-1}
                \left(
                  1 - \frac{ (\hbar^2 / 2 m) [(k_1 - k_{-1})/2]^2 + q B^2}
                               {c_2 [A_1 + (-1)^n A_{-1}]^2}
                \right) \, .
\end{equation}
For a cw solution to exist, $A_0$ must be non-negative, as per
ansatz~(\ref{cw.ansatz}), so its square must be non-negative, $A_0^2
\geq 0$.  The ranges in which the different cw solutions exist depends
on the magnitude and sign of the quantity $[\hbar^2 (k_1 - k_{-1})^2 /
8 m + q B^2] / c_2$.  If the quadratic Zeeman coefficient is
non-negative ($q \geq 0$), the sign of this is either zero or equal to
the sign of $c_2$.  If the quadratic Zeeman coefficient is negative
($q < 0$), a sufficiently strong magnetic field can change the sign of
the quantity.  For particles with $F=1$, the quadratic Zeeman
coefficient is normally positive, $q > 0$.  (Particles with $F=2$,
normally have negative quadratic Zeeman coefficients, $q < 0$.)  The
sign and magnitude of $q$ can be changed, however, by using the
alternating current Stark shift with microwave radiation
\cite{Leslie.2009, Gerbier.2006, Ueda.2012}.  We will refer to
$[\hbar^2 (k_1 - k_{-1})^2 / 8 m + q B^2] / c_2 >0$ as the generalized
antiferromagnetic case, and $[\hbar^2 (k_1 - k_{-1})^2 / 8 m + q B^2]
/ c_2 < 0$ as the generalized ferromagnetic case.
In the generalized antiferromagnetic case, there are cw solutions with
even $n$ (which we will denote by $n=0$ in the labels below) when
\begin{equation}
    (A_1 + A_{-1})^2 \geq \frac{1}{c_2} \left[ \frac{\hbar^2}{2 m}
    \left( \frac{k_1 - k_{-1}}{2} \right)^2 + q B^2 \right] \, ,
\end{equation}
and there are cw solutions with odd $n$ (which we will denote by
$n=1$) when
\begin{equation}
    0 < (A_1 - A_{-1})^2 \leq \frac{1}{c_2} \left[ \frac{\hbar^2}{2 m}
    \left( \frac{k_1 - k_{-1}}{2} \right)^2 + q B^2 \right] \, .
\end{equation}
There are no $n=1$ type cw solutions when $A_1 = A_{-1}$ because the
difference of the amplitudes in the denominator of
Eq.~(\ref{cw.A0.solution}) makes the value of $A_0$ go to infinity as
$(A_1 - A_{-1})$ approaches zero.
In the generalized ferromagnetic case, there are $n=0$ cw solutions
for all values of the spin $M_F=0$ and spin $M_F=-1$ magnitudes ($A_1,
A_{-1} \in {\cal R}^{\geq 0}$).  The generalized ferromagnetic case
does not support any $n=1$ type cw solutions.
Recall that there exist cw solutions with vanishing $M_F=0$ BEC fields
for any values of the $M_F=\pm 1$ BEC magnitudes, whether the BEC is
ferromagnetic or antiferromagnetic.
This existence ranges are summarized in Table~\ref{Table.existence}
and in Fig.~\ref{Fig.CWs.c2positive}.

\begin{table}[ht]
\begin{center}
\begin{tabular}{| c || c | c | c |}
\hline
$x \equiv [ \hbar^2 (k_1 - k_{-1})^2 / 8 m + q B^2] / c_2$
&
cw (CNLS)
&
cw ($n=0$)
&
cw ($n=1$)
\\  \hline \hline
$> 0$
& 
$A_1, A_{-1} \in {\cal R}^{\geq 0}$
&
$(A_1 + A_{-1})^2 \geq x$
&
$0 < (A_1 - A_{-1})^2 \leq x$ 
\\ \hline
$\leq 0$
&
$A_1, A_{-1} \in {\cal R}^{\geq 0}$
&
$A_1, A_{-1} \in {\cal R}^{\geq 0}$
&
No solutions. \\
\hline
\end{tabular}
\end{center}
\caption{Existence ranges of the three possible cw solutions, for
different values of the spin-dependent nonlinear coefficient $c_2$,
particle mass $m$, quadratic Zeeman coefficient $q$, magnetic field
$B$, and given difference in the wavenumbers between the $M_F = \pm 1$
spin components.  The allowed magnitudes of the amplitudes of the
fields for the BEC with spin $M_F = \pm 1$ are denoted by $A_{\pm 1}$,
as in the ansatz~(\ref{cw.ansatz}), and the magnitude of the
amplitudes of the BEC with spin $M_F = 0$ is either $0$ (``CNLS'') or
goes according to the formula~(\ref{cw.A0.solution}).}
\label{Table.existence} 
\end{table}

The Hamiltonian density of the cw solutions is
\begin{eqnarray}
\label{Hamiltonian.cw}
\mathcal{H} 
& = & \frac{\hbar^2}{2 m} (k_1^2 A_1^2 + k_0^2 A_0^2 + k_{-1}^2 A_{-1}^2) + 
\frac{c_0}{2}(A_1^2 + A_0^2 + A_{-1}^2)^2
\nonumber \\
& + & c_2
[ \frac{1}{2}(A_1^2 - A_{-1}^2)^2 + A_0^2 (A_1^2 + A_{-1}^2 + 
(-1)^n A_1 A_{-1}) ]
+ q B^2 (A_1^2 + A_{-1}^2) \, ,
% + (-p B + q B^2) A_1^2 + (p B + q B^2) A_{-1}^2 ,
\end{eqnarray}
where $A_0$ may either be zero or one of the non-zero cw solutions
[Eq.~(\ref{cw.A0.solution})], and $k_0=(k_1+k_{-1})/2$.

Note that all three families of cw solutions include solutions in 
which the spin components have mutually
different wavenumbers -- the solution with identical wavenumbers is a 
limiting case.  This implies that the different components may
move relative to one another.  In the mathematical model, the domain
is infinite.  In an experiment with cws with different wavenumbers in
the different components, one would have to either ensure that the
fields remain in a specific location for long enough, either by doing
the experiment before the edges encroach on the middle, replenishing
the fields at the edges, or arranging the fields in a ring (confined
by a toroidal potential) \cite{Gupta.2005, Olson.2007,
Lesanovsky.2007, Henderson.2009, Ryu.2007}.

\begin{figure}
\centering
\includegraphics[width=19 cm]{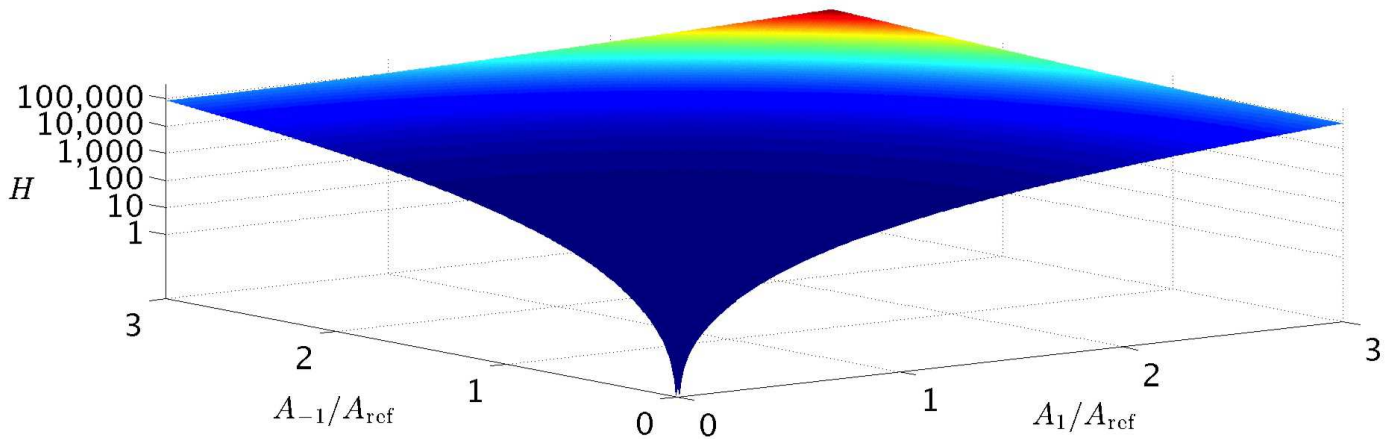}
\caption{(Color online) Hamiltonian density of cw solutions of a BEC
of $^{23}$Na atoms in their ground state, having no particles with
spin component $M_F=0$.  The quantities are shown in dimensionless
units [see Eq.~(\ref{Eq:nondimensionalize})], for generality and to
avoid large amounts of notation.  The cws in the figure have a fixed
unit difference between the $M_F=1$ and $M_F=-1$ spin components
wavenumbers, $k_1 - k_{-1} = 1$, and the magnetic field is zero.  The
Hamiltonian density is shown for a range of magnitudes of the BEC
fields with $M_F=\pm 1$.  The (dimensionless) reference amplitude on
the axes is $A_\mathrm{ref} \equiv [ \hbar^2 (k_1 - k_{-1})^2 / 8 m +
q B^2] / c_2$.}
\label{Fig.Na23.CNLS.H}
\end{figure}

\begin{figure}
\centering
\includegraphics[width=19 cm]{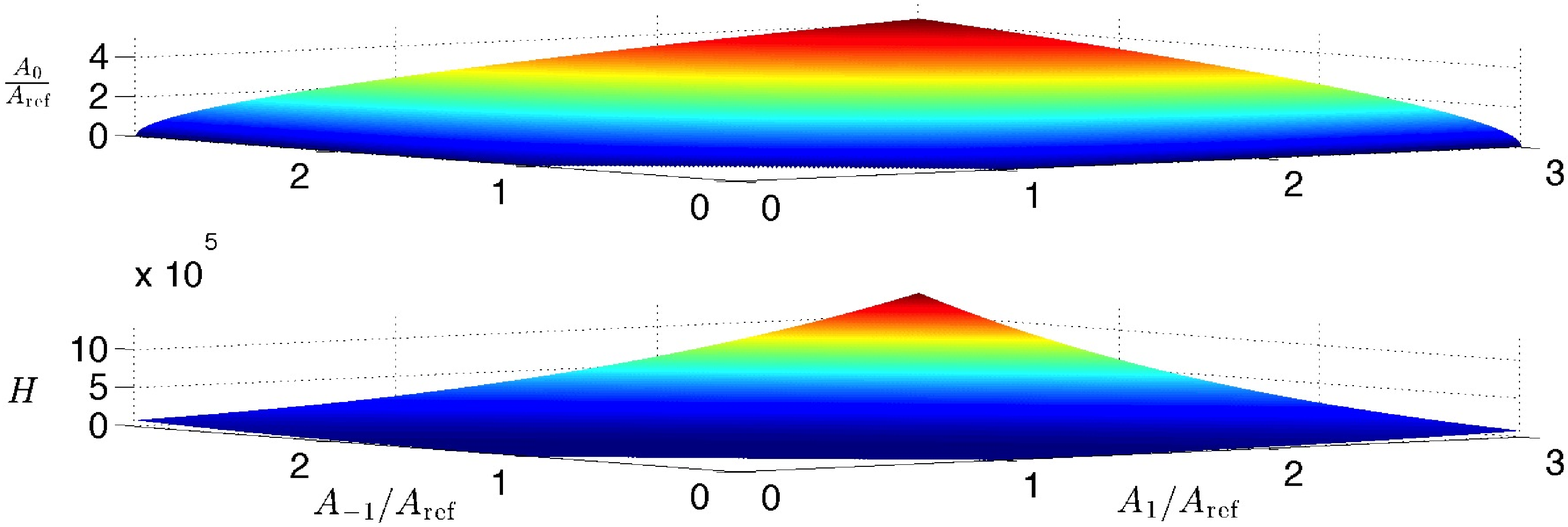}
\caption{(Color online) Amplitude of the spin $M_F=0$ component and
Hamiltonian density for cw solutions of a BEC of $^{23}$Na BEC, with
the relative phase in the spin $M_F=0$ component corresponding to even $n$
in the trial function~(\ref{cw.ansatz}), with $\theta_1 + \theta_{-1}
- 2\theta_0 = n \pi$.  The quantities are shown in dimensionless units
[see Eqs.~(\ref{Eq:nondimensionalize})].  The cws in the figure have a
fixed unit difference between the $M_F=1$ and $M_F=-1$ spin components
wavenumbers, $k_1 - k_{-1} = 1$, and the magnetic field is zero.  The
magnitude of the spin $M_F=0$ amplitude and the Hamiltonian density of the
cw is shown for a range of magnitudes of the BEC fields with $M_F=\pm
1$.  The (dimensionless) reference amplitude is $A_\mathrm{ref} \equiv
[ \hbar^2 (k_1 - k_{-1})^2 / 8 m + q B^2] / c_2$.}
\label{Fig.Na23.n0.A0_H}
\end{figure}

\begin{figure}
\centering
\includegraphics[width=19 cm]{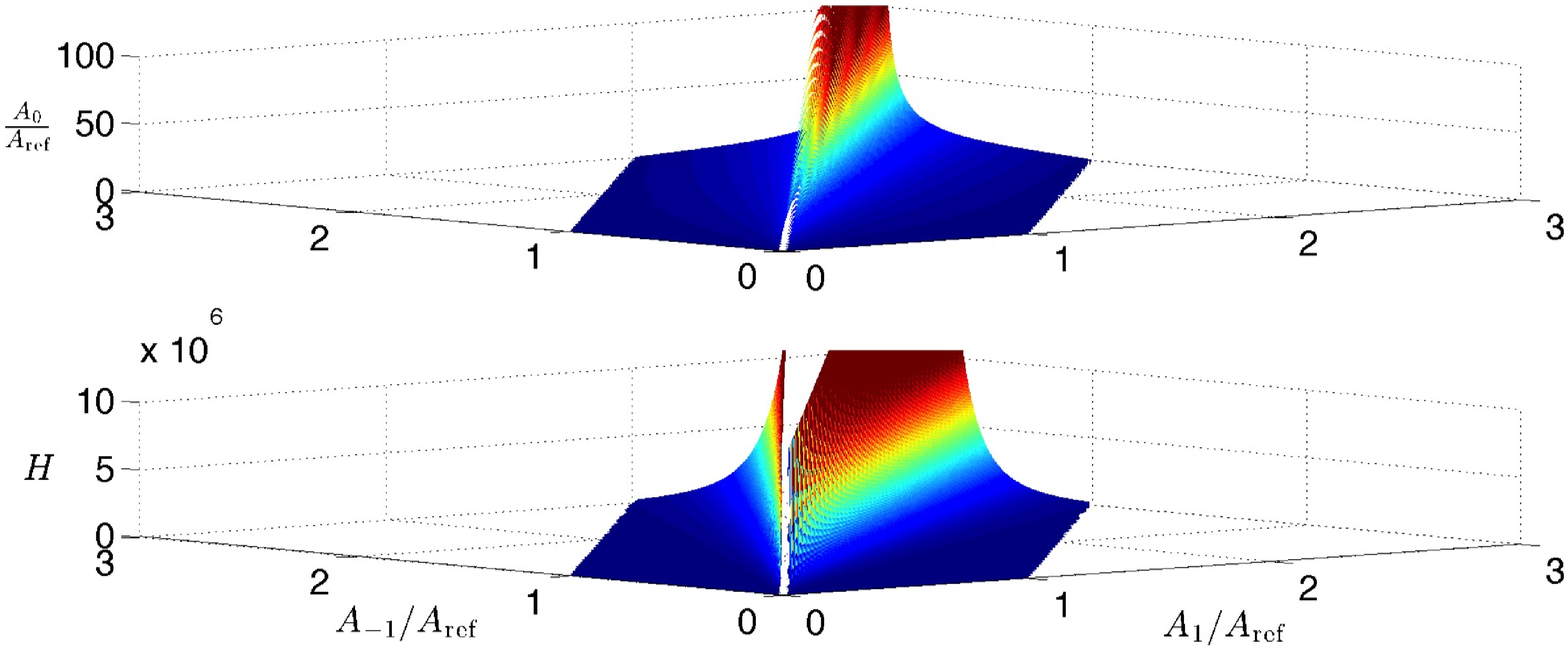}
\caption{(Color online) Amplitude of the spin $M_F=0$ component and
Hamiltonian density for cw solutions of a BEC of $^{23}$Na BEC, with
the relative phase in the spin $M_F=0$ component corresponding to odd $n$
in the trial function Eq.~(\ref{cw.ansatz}), with $\theta_1 +
\theta_{-1} - 2\theta_0 = n \pi$.  The quantities are shown in
dimensionless units [see Eq.~(\ref{Eq:nondimensionalize})].  The cws
in the figure have a fixed unit difference between the $M_F=1$ and
$M_F=-1$ spin components wavenumbers, $k_1 - k_{-1} = 1$, and the
magnetic field is zero.  The magnitude of the spin $M_F=0$ amplitude and
the Hamiltonian density of the cw is shown for a range of magnitudes
of the BEC fields with $M_F=\pm 1$.  The (dimensionless) reference
amplitude is $A_\mathrm{ref} \equiv [ \hbar^2 (k_1 - k_{-1})^2 / 8 m +
q B^2] / c_2$.  For odd $n$, the particle density of the $M_F = 0$ BEC
(as well as its Hamiltonian density) approaches infinity as the values
of the densities of $M_F = 1$ and $M_F = -1$ fields approach
equality.}
\label{Fig.Na23.n1.A0_H}
\end{figure}

\begin{figure}
\centering
\includegraphics[width=19 cm]{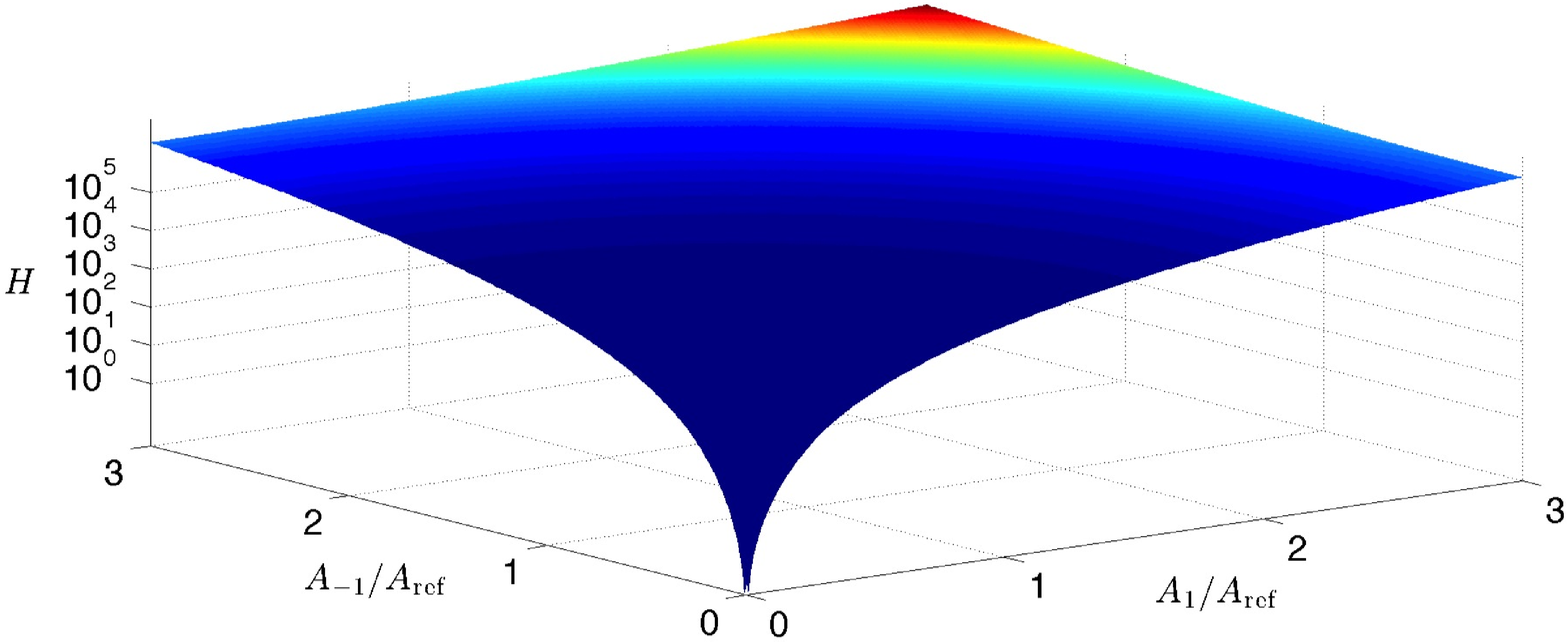}
\caption{(Color online) Hamiltonian density of cw solutions of a BEC
of $^{87}$Rb atoms in their ground state, having no particles with
spin component $M_F=0$.  The quantities are shown in dimensionless
units [see Eq.~(\ref{Eq:nondimensionalize})], for generality and to
avoid large amounts of notation.  The cws have a fixed unit difference
between the $M_F=1$ and $M_F=-1$ spin components wavenumbers, $k_1 -
k_{-1} = 1$, and the magnetic field is zero.  The Hamiltonian density
is shown for a range of magnitudes of the BEC fields with $M_F=\pm 1$.
The (dimensionless) reference amplitude on the axes is $A_\mathrm{ref}
\equiv -[\hbar^2 (k_1 - k_{-1})^2 / 8 m + q B^2] / c_2$.  }
\label{Fig.Rb87.CNLS.H}
\end{figure}

\begin{figure}
\centering
\includegraphics[width=19 cm]{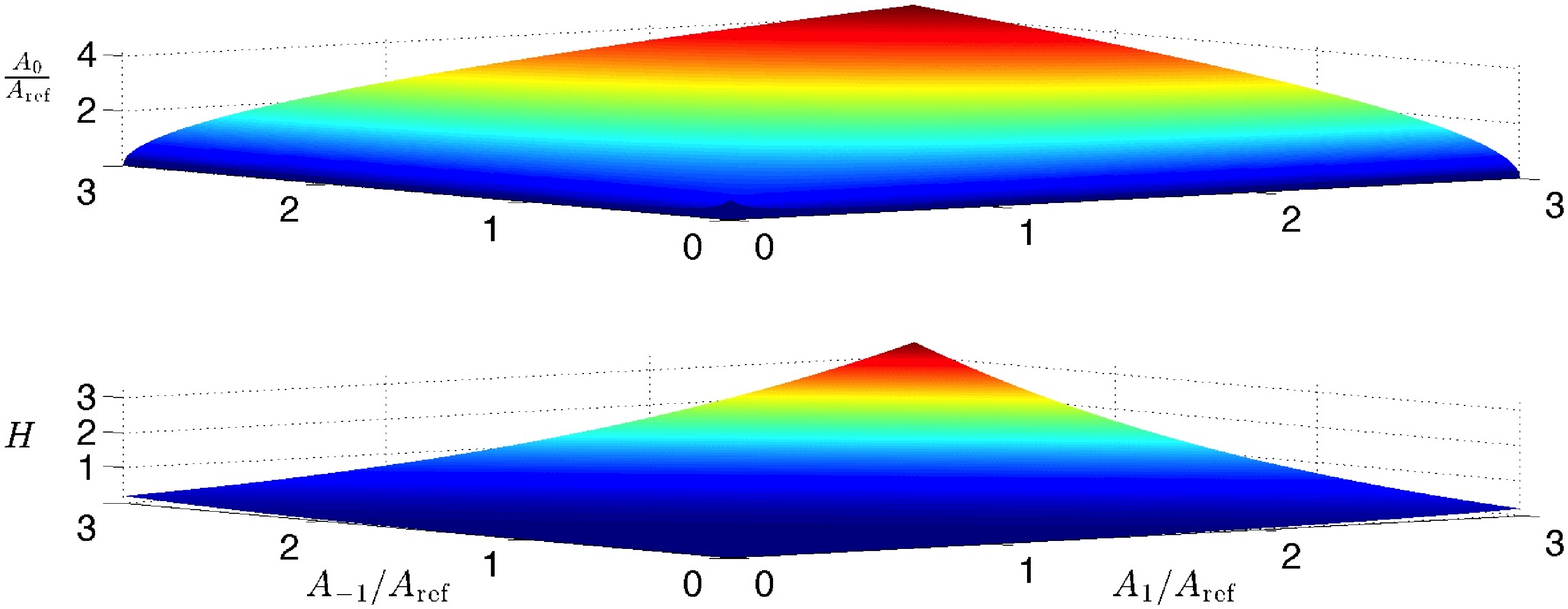}
\caption{(Color online) Amplitude of the spin $M_F=0$ component and
Hamiltonian density for cw solutions of a BEC of $^{87}$Rb BEC, with
the relative phase in the spin $M_F=0$ component corresponding to even $n$
in the trial function Eq.~(\ref{cw.ansatz}), with $\theta_1 +
\theta_{-1} - 2\theta_0 = n \pi$.  The quantities are shown in
dimensionless units [see Eq.~(\ref{Eq:nondimensionalize})].  These cws
have a fixed unit difference between the $M_F=1$ and $M_F=-1$ spin
components wavenumbers, $k_1 - k_{-1} = 1$, and the magnetic field is
zero.  The magnitude of the spin $M_F=0$ amplitude and the Hamiltonian
density of the cw is shown for a range of magnitudes of the BEC fields
with $M_F=\pm 1$.  The (dimensionless) reference amplitude is
$A_\mathrm{ref} \equiv -[\hbar^2 (k_1 - k_{-1})^2 / 8 m + q B^2] /
c_2$.  }
\label{Fig.Rb87.n0.A0_H}
\end{figure}

Figures~\ref{Fig.Na23.CNLS.H} through~\ref{Fig.Rb87.n0.A0_H} show, for
representative cw solutions, the values of the amplitudes of the
spin $M_F=0$ fields and the Hamiltonian densities.  In these sets of
figures, the nonlinear coefficients are those of $^{23}$Na and
$^{87}$Rb, and the the plots are scaled such that they represent any
generalized antiferromagnetic and ferromagnetic cw solutions (though
excluding the sign changes that can come from a negative quadratic
Zeeman coefficients with a non-zero magnetic field).

We carried out a stability analysis of the cw solutions within the
continuous wave assumption, i.e., without looking for modulational
(Benjamin-Feir) instabilities or sound waves (alias Bogoliubov
excitations or phonons), that is, small perturbations of arbitrary
wavenumber on top of the cw solutions
\cite{Bogoliubov.1947, BespalovTalanov.1966, Andrews.1997, Andrews.1998}. 
A study of the sound waves and modulational stabilities is a large and lengthy topic.
We will defer these results to another article.  References~\cite{Robins.2001,
Konotop.2002, Li.2005, KawaguchiUeda.2012} analyze sound waves and
modulational instabilities on top of \textit{subsets} of the complete set of
possible cw solutions herein.  The stability analysis generalizes the
ansatz~(\ref{cw.ansatz}) by allowing the variables for the magnitudes
of the amplitudes $A_{1,0,-1}$, their phases $\theta_{1,0,-1}$, and
wavenumbers $k_{1,0,-1}$ to vary with time.  Inserting this ansatz
into the dynamical equations~(\ref{Eqs:SpinorBEC}) gives the results
that the total particle density, $N \equiv A_1^2 + A_0^2 + A_{-1}^2$,
and the magnetization, $M \equiv A_1^2 - A_{-1}^2$, are conserved.
Moreover, the wavenumbers $k_{1,0,-1}$ are do not change with time,
due to the physics with the cw assumption.  We may reduce the
oscillations about the fixed points that represent the cw solutions to
the dynamics of two variables, $A_0(t)^2$ and $[\theta_1(t) +
\theta_{-1}(t) - 2\theta_0(t)]$ which are conjugate to each other.
The fixed points are stable to small perturbations, with frequencies
\begin{equation}
\label{CW:n0n1.oscillations}
    \omega_{n=0,1}^2 = 4 \left( \frac{c_2}{\hbar} \right)^2
    A_1 A_0^2 A_{-1}
    \left[3 (-1)^n + 2 \left(\frac{A_1}{A_{-1}} + \frac{A_{-1}}{A_1}\right)
    + \frac{(A_1^2 - A_{-1}^2)^2 A_0^2}{4 A_1^3 A_{-1}^3}
    \right] \, .
\end{equation}
The frequencies are always real-valued.  Thus, all the $n=0,1$ cw
solutions are stable within the continuous wave framework (i.e., not
considering modulational instabilities, which due to length is
deferred to another paper).  Physically, these are the frequencies
of oscillatory coherent spin mixing \cite{Zhang.2005, Gerbier.2006, MurPetit.2009}.

This oscillation is not the same as that in
Refs.~\cite{SzankowskiTrippenbachInfeldRowlands.2010}
and\cite{SzankowskiTrippenbachInfeldRowlands.2011}.  In those papers,
the frame of reference can be rotated such that the solutions have no
BEC component with spin $M_F=0$.  The governing equations then reduce
to two coupled nonlinear Schr\"odinger equations, without any
parametric terms that either mix the spin components or lock their
frequencies (chemical potentials) together.
If such solutions are viewed in a rotated
reference frame that mixes the spin components, there is an
oscillation due to the beating of one frequency and wavenumber against
another.
The oscillations with frequencies~(\ref{CW:n0n1.oscillations}), in
contrast, cannot be eliminated by a change in the variables.  There is
just one oscillatory frequency for a given near-cw solution.  One
could rotate such a near-CW solution with a small oscillation into
another reference frame, and get oscillations as in
Refs.~\cite{SzankowskiTrippenbachInfeldRowlands.2010,
SzankowskiTrippenbachInfeldRowlands.2011} \textit{on top of} the
frequency~(\ref{CW:n0n1.oscillations}) of the small variations about
the fixed point.
The oscillations in
Refs.~\cite{SzankowskiTrippenbachInfeldRowlands.2010} can be affected
by linear Zeeman splitting; the oscillation
frequencies~(\ref{CW:n0n1.oscillations}) are independent of it.

\section{Summary and Conclusions}
\label{Sec:SpinorBEC.conclusions}

We obtained the most general continuous wave (plane wave) solutions
for spinor BECs for spin $F=1$, with linear and quadratic Zeeman
splitting due to a magnetic field.  We do not make the assumption that
the wavenumbers of the different spin components are all the same.
The physics only requires that cw solutions have wavenumbers and
frequencies in the $M_F=0$ fields that are the average of the
quantities in the $M_F=\pm 1$ fields.  There are three distinct
families of cw solutions.  The first family are the solutions in which
the $M_F = 0$ spin component vanishes.  The second and third families
of cw solutions have non-zero densities of particles with spin $M_F =
0$.  The second and third families of solutions are distinguished from
each other by the phase of the $M_F=0$ BEC field with respect to the
average phase of the $M_F=\pm 1$ BEC fields.  Other things being
equal, the different phases in the $M_F=0$ fields give the cw
solutions different densities of spin $M_F=0$ particles.  The third
family does not allow all the wavenumbers to be identical; in this
limit, the density of $M_F=0$ particles is asymptotically large.

The first family of solutions, without spin $M_F=0$ particles, is
governed by the coupled nonlinear Schr\"odinger equations, which have
been studied in great detail.  The solutions of the second and third
type depend on whether the BEC is ferromagnetic or polar
(antiferromagnetic).  There are always cw solutions of the first type,
and there is always at least one solution of the second or third type.

The linear Zeeman splitting term, with a constant uniform magnetic
field, can be eliminated by a gauge transformation, so it does not
change the dynamics.  Quadratic Zeeman splitting cannot be eliminated,
and has nontrivial effects.  If the quadratic Zeeman coefficient is
positive (which is the usual case for $F=1$ particles), it's effects
on the cw solutions are similar to the effects of a difference in the
wavenumbers in the spin components.  If the quadratic Zeeman
coefficient is negative (which can be achieved by using microwaves and
the Stark effect), a sufficiently large magnetic field can made the cw
solutions in the $F=1$ spinor BEC switch from ferromagnetic to
antiferromagnetic and \textit{vice versa}.

The cw solutions of the second and third families are stable to small
perturbations within the cw ansatz.  We calculated the frequency of
these perturbations, which correspond to oscillatory coherent spin
mixing.

\begin{acknowledgments}
This work was supported in part by grants from the Israel Science
Foundation (No.~2011295) and the James Franck German-Israel Binational
Program.
\end{acknowledgments}


\begin{thebibliography}{99}

\bibitem{Fried.1998}
D. G. Fried, T. C. Killian, L. Willmann, D. Landhuis, S. C. Moss, 
D. Kleppner, and T. J. Greytak,
``Bose-Einstein Condensation of Atomic Hydrogen,''
Phys. Rev. Lett. \textbf{81}, 3811-3814 (1998).

\bibitem{Bradley.1995}
C. C. Bradley, C. A. Sackett, J. J. Tollett, and R. G. Hulet ,
``Evidence of Bose-Einstein Condensation in an Atomic Gas with Attractive 
Interactions,''
Phys. Rev. Lett. \textbf{75}, 1687-1690 (1995). 

\bibitem{Bradley.1997}  % BEC Li
C. C. Bradley, C. A. Sackett, and R. G. Hulet,
``Bose-Einstein Condensation of Lithium: Observation of Limited Condensate 
Number,''
Phys. Rev. Lett. \textbf{78}, 985-989 (1997).

\bibitem{Simkin.1999}
M. V. Simkin and E. G. D. Cohen,
``Magnetic properties of a Bose-Einstein condensate,''
Phys. Rev. A \textbf{59}, 1528-1532 (1999).

% \bibitem{Kraft.2009}
% S. Kraft, F. Vogt, O. Appel, F. Riehle, and U. Sterr,
% ``Bose-Einstein Condensation of Alkaline Earth Atoms: $^{40}$Ca,''
% Phys. Rev. Lett. \textbf{103}, 130401 (2009).
%
% Ca-40 is spin 0 in the ground state. Spinor in excited states.

\bibitem{Roati.2007}
G. Roati, M. Zaccanti, C. D'Errico, J. Catani, M. Modugno, A. Simoni,
M. Inguscio, and G. Modugno,
``$^{39}$K Bose-Einstein Condensate with Tunable Interactions,''
Phys. Rev. Lett. \textbf{99}, 010403 (2007).

\bibitem{Modugno.2001}
G. Modugno, G. Ferrari, G. Roati, R. J. Brecha, A. Simoni, and M. Inguscio,
``Bose-Einstein Condensation of Potassium Atoms by Sympathetic Cooling,''
Science \textbf{294}, 1320-1322 (2001).

\bibitem{Griesmaier.2005}
A. Griesmaier, J. Werner, S. Hensler, J. Stuhler, and T. Pfau,
``Bose-Einstein Condensation of Chromium,''
Phys. Rev. Lett. \textbf{94}, 160401 (2005).

\bibitem{Beaufils.2008}
Q. Beaufils, R. Chicireanu, T. Zanon, B. Laburthe-Tolra, E. Marechal,
L.  Vernac, J.-C. Keller, and O. Gorceix,
``All-optical production of chromium Bose-Einstein condensates,''
Phys. Rev. A \textbf{77}, 061601(R) (2008).

% \bibitem{Stellmer.2009}
% S. Stellmer, M. K. Tey, B. Huang, R. Grimm, and F. Schreck,
% ``Bose-Einstein Condensation of Strontium,''
% Phys. Rev. Lett. \textbf{103}, 200401 (2009).
%
% Sr-84 is spin 0 in the ground state.

\bibitem{Barrett.2001}
M. D. Barrett, J. A. Sauer, and M. S. Chapman,
``All-Optical Formation of an Atomic Bose-Einstein Condensate,''
Phys. Rev. Lett. \textbf{87}, 010404 (2001).

\bibitem{Chang.2004}
M.-S. Chang, C. D. Hamley, M. D. Barrett, J. A. Sauer, K. M. Fortier, W. 
Zhang, L. You, and M. S. Chapman,
``Observation of Spinor Dynamics in Optically Trapped $^{87}$Rb
Bose-Einstein Condensates,''
Phys. Rev. Lett. \textbf{92}, 140403 (2004).

\bibitem{Sadler.2006}
L. E. Sadler, J. M. Higbie, S. R. Leslie, M. Vengalattore, and
D. M. Stamper-Kurn,
``Spontaneous symmetry breaking in a quenched ferromagnetic spinor 
Bose-Einstein condensate,''
Nature (London) \textbf{443}, 312-315 (2006).

\bibitem{Leslie.2009}
S. R. Leslie, J. Guzman, M. Vengalattore, J. D. Sau, M. L. Cohen, and
D. M. Stamper-Kurn,
``Amplification of fluctuations in a spinor Bose-Einstein condensate,''
Phys. Rev. A \textbf{79}, 043631 (2009).

\bibitem{Weber.2003}
T. Weber, J. Herbig, M. Mark, H.-C. Nagerl, and R. Grimm,
``Bose-Einstein Condensation of Cesium,''
Science \textbf{299}, 232-235 (2003).

\bibitem{Lu.2011}
M. Lu, N. Q. Burdick, S. H. Youn, and B. L. Lev,
``Strongly Dipolar Bose-Einstein Condensate of Dysporium,''
Phys. Rev. Lett. \textbf{107}, 190401 (2011).

\bibitem{Aikawa.2012}
K. Aikawa, A. Frisch, M. Mark, S. Baier, A. Rietzler, R. Grimm, and
F. Ferlaino,
``Bose-Einstein Condensation of Erbium,''
Phys. Rev. Lett. \textbf{108}, 210401 (2012).

% \bibitem{Fukuhara.2007}
% T. Fukuhara, S. Sugawa, and Y. Takahashi,
% ``Bose-Einstein condensation of an ytterbium isotope,''
% Phys. Rev. A \textbf{76}, 051604 (2007).
%
% Yb-170 is spin 0 in the ground state. Spinor in excited states.

% \bibitem{Takasu.2003}
% Y. Takasu, K. Maki, K. Komori, T. Takano, K. Honda, M. Kumakura, T. Yabuzaki, 
% and Y. Takahashi,
% ``Spin-Singlet Bose-Einstein Condensation of Two-Electron Atoms,''
% Phys. Rev. Lett. \textbf{91}, 040404 (2003).
% 
% Yb-174 is spin 0 in the ground state. Spinor in excited states.

% Toroidal potentials:
% Gupta.2005, Olson.2007, Lesanovsky.2007, Henderson.2009, Ryu.2007
\bibitem{Gupta.2005}
S. Gupta, K. W. Murch, K. L. Moore, T. P. Purdy, and D. M. Stamper-Kurn,
``Bose-Einstein Condensation in a Circular Waveguide,''
Phys. Rev. Lett. \textbf{95}, 143201 (2005).

\bibitem{Olson.2007} % N. B. Optical trap 
S. E. Olson, M. L. Terraciano, M. Bashkansky, and F. K. Fatemi,
``Cold-atom confinement in an all-optical dark ring trap,''
Phys. Rev. A \textbf{76}, 061404(R) (2007).

\bibitem{Lesanovsky.2007}
I. Lesanovsky and W. von Klitzing,
``Time-Averaged Adiabatic Potentials: Versatile Matter-Wave Guides and Atom 
Traps,''
Phys. Rev. Lett. \textbf{99}, 083001 (2007).

\bibitem{Henderson.2009}
K. Henderson, C. Ryu, C. MacCormick, and M. G. Boshier,
``Experimental demonstration of painting arbitrary and dynamic potentials
for Bose-Einstein condensates,''
New J. Phys. \textbf{11}, 043030 (2009).

\bibitem{Ryu.2007}
C. Ryu, M. F. Andersen, P. Clade, V. Natarajan, K. Helmerson, and
W. D. Phillips,
``Observation of Persistent Flow of a Bose-Einstein Condensate
in a Toroidal Trap,''
Phys. Rev. Lett. \textbf{99}, 260401 (2007).

\bibitem{Ueda.2012}
M. Ueda,
``Bose Gases with Nonzero Spin,''
Ann. Rev. Condens. Matter Phys. \textbf{3}, 263-283 (2012).

\bibitem{KawaguchiUeda.2012}
Y. Kawaguchi and M. Ueda,
``Spinor Bose-Einsetin Condensates,''
Physics Reports \textbf{520}, 253-381 (2012).

\bibitem{Saito.2007}  % Zeeman
H. Saito, Y. Kawaguchi, and M. Ueda,
``Topological defect formation in a quenched ferromagnetic Bose-Einstein 
condensates,''  %(sic -- the plural)
Phys. Rev. A \textbf{75}, 013621 (2007).

\bibitem{Lamacraft.2007}  % Zeeman
A. Lamacraft,
``Quantum Quenches in a Spinor Condensate,''
Phys. Rev. Lett. \textbf{98}, 160404 (2007).

\bibitem{Zhang.2005}  % Zeeman
W. Zhang, D. L. Zhou, M.-S. Chang, M. S. Chapman, and L. You,
``Coherent spin mixing dynamics in a spin-1 atomic condensate,''
Phys. Rev. A \textbf{72}, 013602 (2005).

\bibitem{Gerbier.2006}
F. Gerbier, A. Widera, S. F\"{o}lling, O. Mandel, and I. Bloch,
``Resonant control of spin dynamics in ultracold quantum gases by 
microwave dressing,''
Phys. Rev. A \textbf{73}, 041602(R) (2006).

\bibitem{Robins.2001}
N. P. Robins, W. Zhang, E. A. Ostrovskaya, and Y. S. Kivshar,
``Modulational instability of spinor condensates,''
Phys. Rev. A \textbf{64}, 021601(R) (2001).

\bibitem{Konotop.2002}
V. V. Konotop and M. Salerno,
``Modulational instability in Bose-Einstein condensates in optical lattices,''
Phys. Rev. A \textbf{65}, 021602(R) (2002).

\bibitem{Li.2005} % Just a little MI, numerical
L. Li, Z. Li, B. A. Malomed, D. Mihalache, and W. M. Liu,
``Exact soliton solutions and nonlinear modulation instability in spinor 
Bose-Einstein condensates,''
Phys. Rev. A \textbf{72}, 033611 (2005).

\bibitem{IedaMiyakawaWadati.2004}
J. Ieda, T. Miyakawa, and M. Wadati,
``Exact Analysis of Soliton Dynamics in Spinor Bose-Einstein Condensates,''
Phys. Rev. Lett. \textbf{93}, 194102 (2004).

\bibitem{WadatiTsuchida.2006}
M. Wadati and N. Tsuchida,
``Wave Propagations in the F=1 Spinor Bose-Einstein Condensates,''
J. Phys. Soc. Jpn. \textbf{75}, 014301 (2006).

\bibitem{UchiyamaIedaWadati.2006}
M. Uchiyama, J. Ieda, and M. Wadati,
``Dark solitons in F=1 spinor Bose-Einstein condensate,''
 J. Phys. Soc. Jpn. \textbf{75}, 064002 (2006).

\bibitem{MurPetit.2009}
J. Mur-Petit,
``Spin dynamics and structure formation in a spin-1 condensate in a magnetic field,"
Phys. Rev. A \textbf{79}, 063603 (2009)

\bibitem{SzankowskiTrippenbachInfeldRowlands.2010}
P. Szankowski, M. Trippenbach, and E. Infeld, and G. Rowlands,
``Oscillating Solitons in a Three-Component Bose-Einstein Condensate,''
Phys. Rev. Lett \textbf{105}, 125302 (2010).

\bibitem{SzankowskiTrippenbachInfeldRowlands.2011}
P. Szankowski, M. Trippenbach, E. Infeld, and G. Rowlands,
``Class of compact entities in three-component Bose-Einstein condensates,''
Phys. Rev. A \textbf{83}, 013626 (2011).

\bibitem{OhmiMachida.1998}
T. Ohmi and K. Machida,
``Bose-Einstein Condensation with Internal Degrees of Freedom in Alkali Atom Gases,''
J. Phys. Soc. Jpn. \textbf{67}, 1822-1825 (1998).

\bibitem{Ho.1998}
T. -L. Ho,
``Spinor Bose Condensates in Optical Traps,''
Phys. Rev. Lett. \textbf{81}, 742-745 (1998).

\bibitem{Olshanii.1998}
M. Olshanii,
``Atomic Scattering in the Presence of an External Confinement
and a Gas of Impenetrable Bosons,''
Phys. Rev. Lett. \textbf{81}, 938 (1998).
 
\bibitem{Bergeman.2003}
T. Bergeman, M. G. Moore, and M. Olshanii,
``Atom-Atom Scattering under Cylindrical Harmonic Confinement:
Numerical and Analytic Studies of the Confinement Induced Resonance,''
Phys. Rev. Lett. \textbf{91}, 163201 (2003).

 \bibitem{Chang.2005}
M.-S. Chang, Q. Qin, W. Zhang, L. You, and M. S. Chapman,
``Coherent spinor dynamics in a spin-1 Bose condensate,''
Nature Phys. \textbf{1}, 111-116 (2005).

\bibitem{KlausenBohnGreene.2001}
N. N. Klausen, J. L. Bohn, and C. H. Greene,
``Nature of spinor Bose-Einstein condensates in rubidium,''
Phys. Rev. A \textbf{64}, 053602 (2001).

\bibitem{KempenKokkelmansHeinzenVerhaar.2002}
E. G. M. van Kempen, S. J. J. M. F. Kokkelmans, D. J. Heinzen, and 
B. J. Verhaar,
``Interisotope Determination of Ultracold Rubidium Interactions 
from Three High-Precision Experiments,''
Phys. Rev. Lett. \textbf{88}, 093201 (2002).

\bibitem{Manakov.1973}
S. V. Manakov,
``On the theory of two-dimensional stationary
self-focusing of electromagnetic waves,''
Zh. Eksp. Teor. Fiz. \textbf{65}, 505-516 (1973)
[Sov. Phys.-JETP \textbf{38}, 248-253 (1974)].

\bibitem{Nakkeeran.1998}
K. Nakkeeran, K. Porsezian, P. S. Sundaram, and A. Mahalingam,
``Optical Solitons in N-Coupled Higher Order Nonlinear Schr\"{o}dinger 
Equations,''
Phys. Rev. Lett. \textbf{80}, 1425-1428 (1998).

\bibitem{IedaWadati.2007}
J. Ieda and M. Wadati,
``Nonlinear Dynamics of Spin Structure in Confined Bose-Einstein 
Condensates,''
J. Low Temp. Phys. \textbf{148}, 405-410 (2007).

\bibitem{ZakharovShabat.1971}
V. E. Zakharov and A. B. Shabat,
``Exact Theory of Two-dimensional Self-focusing and
One-dimensional Self-modulation of Waves in Nonlinear Media,"
Zh. Eksp. Teor. Fiz. \textbf{61}, 118-134 (1971)
[Sov. Phys. JETP \textbf{34}, 62-69 (1972)]. 

\bibitem{ZakharovShabat.1973}
V. E. Zakharov and A. B. Shabat,
``Interaction between solitons in a stable medium,"
Zh. Eksp. Teor. Fiz. \textbf{64}, 1627-1639 (1973)
[Sov. Phys. JETP \textbf{37}, 823-828 (1973)]. 

\bibitem{Yang.2010}
J. Yang,
\textit{Nonlinear Waves in Integrable and Nonintegrable Systems}
(SIAM, Philadelphia, 2010).

\bibitem{ZakharovSchulman.1982}
V. E. Zakharov and E. I. Schulman,
``To the integrability of the system of two coupled nonlinear Schr\"{o}dinger 
equations,''
Physica D \textbf{4}, 270-274 (1982).

\bibitem{Agrawal.2001}
G. P. Agrawal,
\textit{Applications of Nonlinear Fiber Optics}
(Academic, NY, 2001).

\bibitem{KivsharTuritsyn.1993} 
Y. S. Kivshar and S. K. Turitsyn 
``Vector dark solitons,''
Opt. Lett. \textbf{18}, 337-339 (1993).

\bibitem{RadhakrishnanLakshmanan.1995}
R. Radhakrishnan and M. Lakshmanan,
``Bright and dark soliton solutions to coupled nonlinear Schr\"odinger 
equations,''
J. Phys. A: Math. Gen. \textbf{28}, 2683-2692 (1995).

\bibitem{RadhakrishnanLakshmananDaniel.1995}
R. Radhakrishnan, M. Lakshmanan, and M. Daniel,
``Bright and dark optical solitons in coupled higher-order
nonIinear Schr\"odinger equations through singularity 
structure analysis,''
J. Phys. A Math. Gen. \textbf{28}, 7299-7314 (1995).

\bibitem{HaeltermanSheppard.1994}  % Domain wall
M. Haelterman and A. P. Sheppard,
``Bifurcations of the dark soliton and polarization domain walls in 
nonlinear dispersive media,''
Phys. Rev. E \textbf{49}, 4512-4518 (1994).

\bibitem{Malomed.1994}
B. A. Malomed,
``Domain wall between traveling waves,''
Phys. Rev. E \textbf{50}, R3310-R3313 (1994).

% \bibitem{Menyuk.1987} % MI, normal dispersion, birefringence
% C. R. Menyuk,
% ``Nonlinear Pulse Propagation in Birefringent  Optical  Fiber,''
% IEEE J. Quantum Electron. \textbf{QE-23}, 174-176 (1987). 
% parametric-like terms for nearly radially symmetric fibers are
% treated carefully, then dropped

% \bibitem{Agrawal.1987} % MI, normal dispersion, birefringence
% G. P. Agrawal,
% ``Modulation Instability Induced by Cross-Phase Modulation,''
% Phys. Rev. Lett. \textbf{59}, 880-883 (1987).

% \bibitem{LiPromislow.2000}
% Y. A. Li and K. Promislow,
% ``The mechanism of the polarization mode instability in birefringent 
% fiber optics,''
% SIAM J. Math. Anal. \textbf{31}, 1351-1373 (2000).

% \bibitem{Rothenberg.1990} %Experimental MI observation, normal dispersion, birefringence
% J. E. Rothenberg,
% ``Modulational instability for normal dispersion,''
% Phys. Rev. A \textbf{42}, 682-685 (1990).

% \bibitem{KadioBand.2006} 
% D. Kadio and Y. B. Band,
% ``Analysis of a magnetically trapped atom clock,''
% Phys. Rev. A \textbf{74}, 053609 (2006).

\bibitem{Bogoliubov.1947}
N. N. Bogoliubov,
``On the Theory of Superfluidity,''
Izv. Akad. Nauk. SSSR Ser. Fiz. \textbf{11}, 77-90 (1947)
[J. Phys. (Moscow) \textbf{11}, 23-32 (1947)].

\bibitem{BespalovTalanov.1966}
V. I. Bespalov and V. I. Talanov,
``Filamentary Structure of Light Beams in Nonlinear Media,"
Pis'ma Zh. Eksp. Teor. Fiz. \textbf{3}, 471-476 (1966)
[JETP Lett. \textbf{3}, 307-310 (1966)].

\bibitem{Andrews.1997}
M. R. Andrews, D. M. Kurn, H.-J. Miesner, D. S. Durfee, C. G. Townsend,
S. Inouye, and W. Ketterle,
``Propagation of Sound in a Bose-Einstein Condensate,"
Phys. Rev. Lett. \textbf{79}, 553-556 (1997).

\bibitem{Andrews.1998}
M. R. Andrews, D. M. Stamper-Kurn, H.-J. Miesner, D. S. Durfee,
C. G. Townsend, S. Inouye, and W. Ketterle,
``Erratum: Propagation of Sound in a Bose-Einstein Condensate,"
Phys. Rev. Lett. \textbf{80}, 2967 (1998).

\end{thebibliography}
\end{document}